\pdfoutput=1

\documentclass[11pt]{article}

\usepackage[final]{acl}

\usepackage{times}
\usepackage{latexsym}
\usepackage{multicol}
\usepackage{multirow}
\usepackage{booktabs}
\usepackage[T1]{fontenc}

\usepackage[utf8]{inputenc}

\usepackage{microtype}

\usepackage{inconsolata}

\usepackage{graphicx}

\title{Multi-Value-Product Retrieval-Augmented Generation\\
for Industrial Product Attribute Value Identification}

\author{
 \textbf{Huike Zou\textsuperscript{1}},
 \textbf{Haiyang Yang\textsuperscript{1}},
 \textbf{Yindu Su\textsuperscript{2}},
 \textbf{Liyu Chen\textsuperscript{1}},
 \\
 \textbf{Chengbao Lian\textsuperscript{1}},
 \textbf{Qingheng Zhang\textsuperscript{1}}
 \textbf{Shuguang Han\textsuperscript{1,}\footnotemark[2]},
 \textbf{Jufeng Chen\textsuperscript{1}}
\\
 \textsuperscript{1}Xianyu of Alibaba \quad
 \textsuperscript{2}Xiaohongshu Inc.
\\
\textbf{\{zouhuike.zhk,qingheng.zqh,shuguang.sh\}@alibaba-inc.com}
}

\begin{document}
\maketitle
\renewcommand{\thefootnote}{\fnsymbol{footnote}}
\footnotetext[2]{Corresponding author.}
\begin{abstract}
Identifying attribute values from product profiles is a key task for improving product search, recommendation, and business analytics on e-commerce platforms, which we called Product Attribute Value Identification (PAVI) . However, existing PAVI methods face critical challenges, such as cascading errors, inability to handle out-of-distribution (OOD) attribute values, and lack of generalization capability. To address these limitations, we introduce Multi-Value-Product Retrieval-Augmented Generation (MVP-RAG), combining the strengths of retrieval, generation, and classification paradigms. MVP-RAG defines PAVI as a retrieval-generation task, where the product title description serves as the query, and products and attribute values act as the corpus. It first retrieves similar products of the same category and candidate attribute values, and then generates the standardized attribute values. The key advantages of this work are: (1) the proposal of a multi-level retrieval scheme, with products and attribute values as distinct hierarchical levels in PAVI domain (2) attribute value generation of large language model to significantly alleviate the OOD problem and (3) its successful deployment in a real-world industrial environment. Extensive experimental results demonstrate that MVP-RAG performs better than the state-of-the-art baselines.
\end{abstract}

\section{Introduction}
Product attribute values are fundamental components in the e-commerce sector, serving as critical elements within the online retail ecosystem. From the perspective of a seller’s business strategy, meticulously curated product attribute values can significantly enhance traffic acquisition, increase product visibility, and ultimately drive transaction conversion. From the operational standpoint of e-commerce platforms, these attribute values provide fundamental data support for core functionalities, such as optimizing product display algorithms\cite{chen2024ipl}, constructing personalized recommendation systems\cite{truong2022ampsum}, and enabling intelligent question-answering services\cite{gao2019product}, which are essential for improving user experience\cite{vashishth2024enhancing} and platform operational efficiency.
\begin{figure}[t]
  \includegraphics[width=\columnwidth]{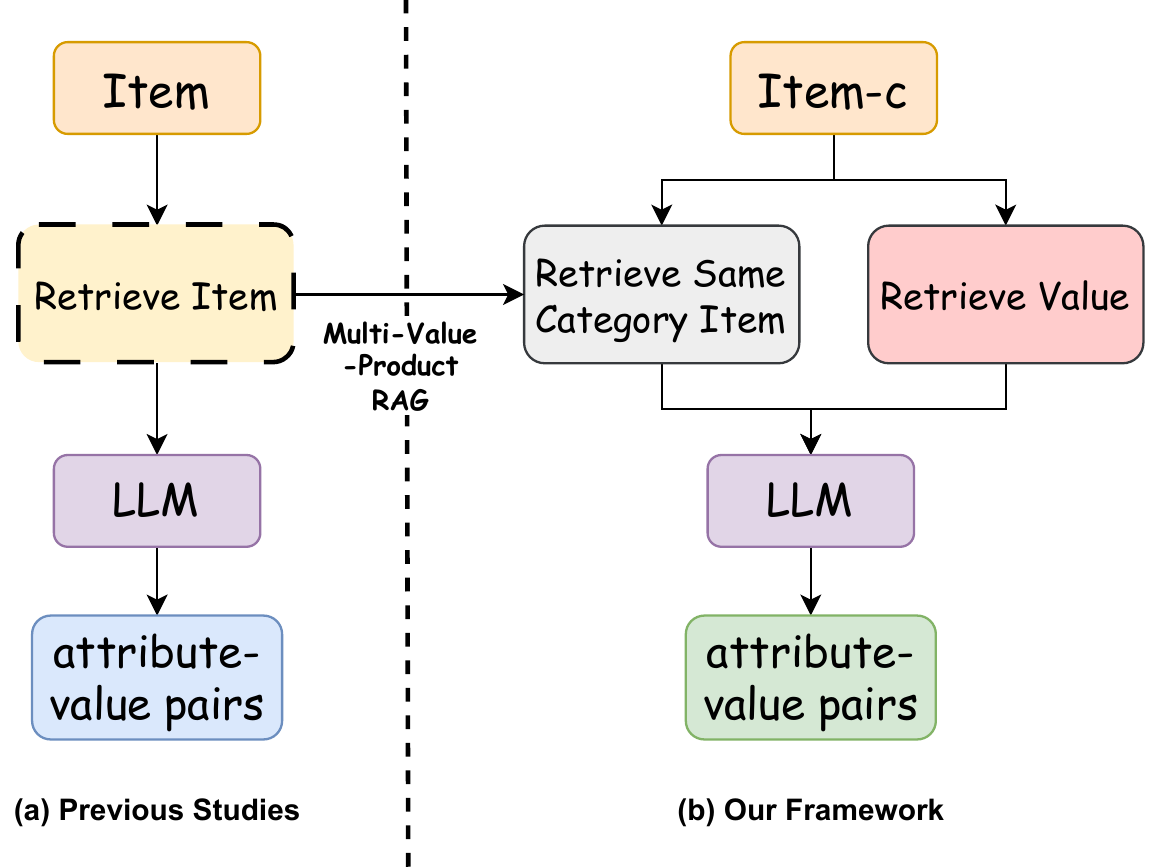}
  \caption{(a) shows the previous LLM and RAG methods, which either retrieve products from the product pool or directly generate attribute values. In contrast, (b) depicts our approach, which retrieves from both product and attribute value perspectives. This not only reduces the hallucination error caused by retrieving irrelevant products, but also ensures the stability of the results. Compared to previous studies, the core of our approach lies in the change in Retrieval.}
  \label{fig:intro}
\end{figure}

Professional sellers are usually able to standardize the selection of product attribute values. However, on second-hand trading platforms like Xianyu\footnote{Xianyu is a C2C e-commerce platform in Alibaba}, individual sellers often struggle to provide complete and accurate product attributes, which severely impacts the circulation efficiency of products. Therefore, achieving automated and precise product attribute value identification (PAVI)\cite{shinzato-etal-2023-unified} is of great significance. Currently, PAVI models can be broadly categorized into two-stage\cite{putthividhya-hu-2011-bootstrapped,Zhang_2021} and one-stage paradigms. 
The two-stage methods involve two steps: product attribute value extraction (PAVE) and alignment. They first extract non-standard attribute values from text using entity recognition\cite{10.1145/3219819.3219839,xu-etal-2019-scaling,yan-etal-2021-adatag} or question-answering\cite{wangqifan,yang-etal-2023-mixpave} techniques, then map them to standard attribute value. The one-stage methods can be further divided into three strategies: classification-based, generation-based, and retrieval-based. The classification-based approach\cite{chen-etal-2022-extreme} treats each attribute value as an independent class for multi-label classification. The generation-based approach\cite{sabeh2024empiricalcomparisongenerativeapproaches,nikolakopoulos2023sagestructuredattributevalue,shinzato-etal-2023-unified} directly generates standard attribute values based on product descriptions. The retrieval-based approach \cite{su2025taclr} encodes product descriptions and attribute values into vector representations, and then selects the relevant attribute values through similarity calculations.
\begin{figure}[t]
\centering
  \includegraphics[width=0.8\columnwidth]{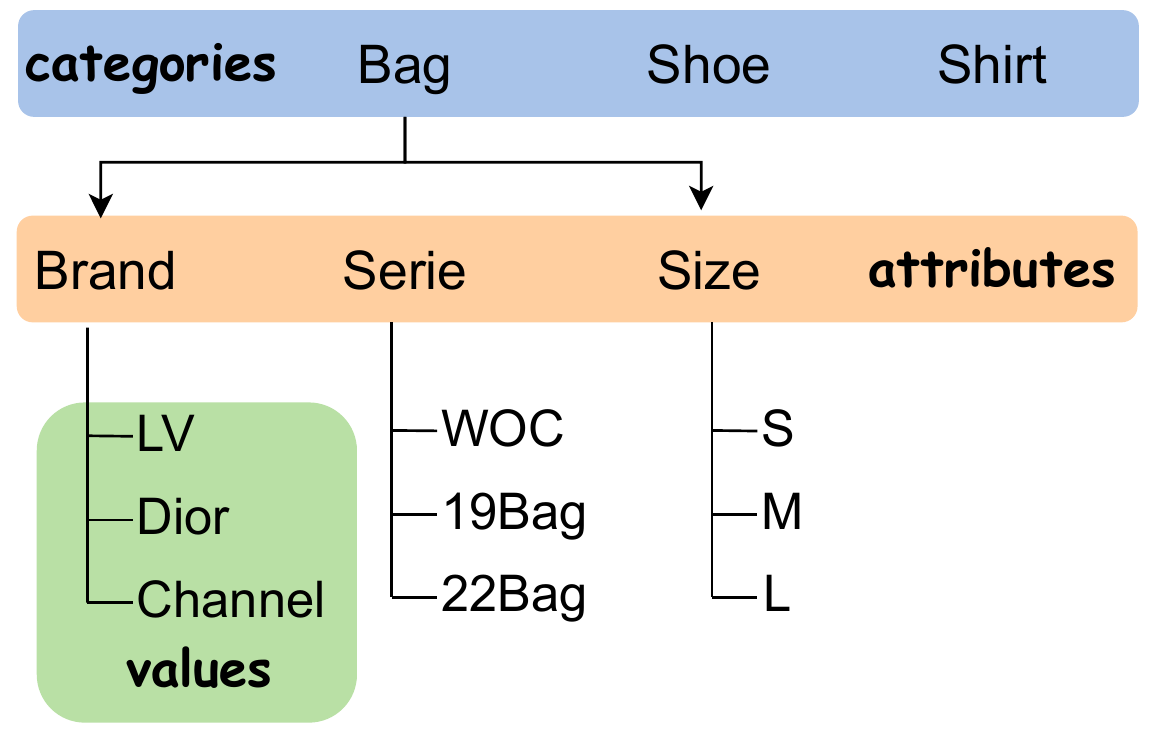}
  \caption{An illustration of a portion of the attribute taxonomy. Each category, such as \textit{Bag}, is linked to multiple attributes, including \textit{Brand}, \textit{Serie}, and \textit{Size}, with standardized values enumerated for each attribute (e.g., \textit{LV}, \textit{Dior}, and \textit{Channel} for \textit{Brand}).}
  \label{fig:taxonomy}
\end{figure}

The existing PAVI methods have their own strengths and weaknesses, and face various technical challenges. The two-stage PAVI methods first extract non-standard attribute values and then align them to the standard attribute value. This approach suffers from cascading errors, where attributes that cannot be identified in the first stage will be completely lost in the second stage. The one-stage methods can address this problem but introduce other issues. The classification-based methods treat attribute values as independent classes, which is a simple approach but is limited in its ability to identify OOD attribute values, making it unsuitable for evolving e-commerce platforms. The generation-based methods view PAVI as an end-to-end task, which can solve the OOD problem, but suffer from the inability to control the output. The retrieval-based methods consider PAVI as a matching task between product and attribute values, but face issues with undefined thresholds and insensitivity to new attribute values. In summary, the existing methods face various problems, including implicit value identification, OOD recognition, and generalization capability. 

In response to the limitations of existing methods, this paper proposes a novel \textbf{M}ulti-\textbf{V}alue-\textbf{P}roduct \textbf{R}etrieval \textbf{A}ugmented \textbf{G}eneration (\textbf{MVP-RAG}) approach, which innovatively combines the strengths of retrieval, generation, and classification paradigms, as shown in Fig \ref{fig:intro}. Our method defines PAVI as a retrieval-generation task: the product title description serves as the query, while the product pool and attribute taxonomy act as the corpus. We first retrieve similar products with identical category and attribute values, and then generate the standardized attribute values. For attribute value retrieval, we leverage TACLR\cite{su2025taclr} to select the top-K attribute values with the highest prediction scores as the candidate set. For product retrieval, we use the BGE model\cite{xiao2024c} to generate vector representations of the product corpus, and then retrieve similar products based on similarity. Based on the retrieval results, we use a large language model (LLM)\cite{zhao2025surveylargelanguagemodels} to generate the attribute values. Additionally, we construct a separate batch of out-of-distribution attribute values and incorporate them into the model training, to enhance the model's ability to discover and predict unknown attribute values.

Our contributions are threefold: (1) We introduce a Multi-Value-Product Retrieval-Augmented Generation (MVP-RAG) method for the PAVI task.
This approach addresses the limitations of existing PAVI methods, which often suffer from cascading errors, inability to handle out-of-distribution (OOD) attribute values, and lack of generalization capability. (2) We incorporate multi-retrieval techniques into PAVI framework, using TACLR for attribute value retrieval and universal representation models for product retrieval. The retrieved attribute values and product information provide valuable contextual cues to guide the subsequent attribute value generation. (3) We validate the effectiveness of MVP-RAG through extensive experiments on proprietary datasets. In addition, MVP-RAG has been successfully deployed in a real-world industrial environment.
\begin{figure*}[t]
\centering
  \includegraphics[width=1.8\columnwidth]{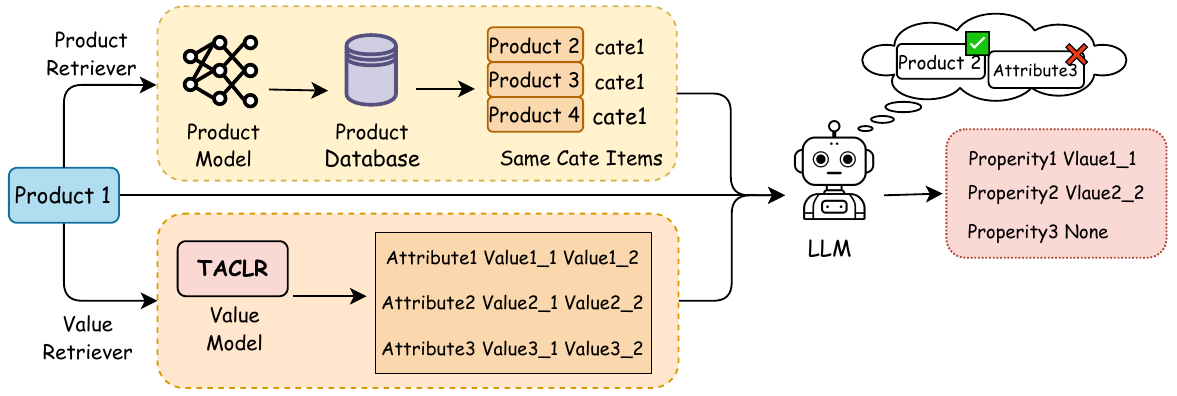}
  \caption{Pipeline of MVP-RAG}
  \label{fig:pipeline}
\end{figure*}

\section{Approach}
As shown in Fig\ref{fig:pipeline}, the overall workflow of MVP-RAG can be divided into two main components: multi-level attribute value-product retrieval (\$\ref{sec1}) and attribute value generation (\$\ref{sec2}). 

\subsection{Multi-level Attribute Value-Product Retrieval}
\label{sec1}
The main purpose of the multi-level attribute value-product retrieval in MVP-RAG is to provide candidate attribute values for the subsequent attribute value generation, as well as to offer product-level few-shot information to guide the selection and generation process, as shown in Fig\ref{fig:pipeline}

\subsubsection{Attribute Value Retrieval}
In the standard retrieval task\cite{schutze2008introduction}, given a query, the goal is to retrieve a set of relevant documents from the corpus. Correspondingly, we treat product information as the query and the standard attribute values as the corpus, and retrieve the most relevant attribute values based on the product information. TACLR\cite{su2025taclr} is the current state-of-the-art approach for the PAVI task based on retrieval, and we follow its setup. For each product, we concatenate the product title t and description d into a sentence in the format \textit{title description} as the input. For each attribute value, we fully utilize the attribute tree information and construct a prompt in the format \textit{a category with attribute being value} as the corpus input.

After constructing the product queries, we leverage the TACLR encoder to obtain the vector representations. Then, for each product query, we calculate the similarity with the attribute values under its corresponding category, and select the top-k candidates as the results.
\subsubsection{Product Retrieval}
Attribute value retrieval can provide a candidate set for attribute value generation, but accurately selecting the most appropriate attribute value from these candidates remains a challenge. Product-level few-shot examples can provide reference cases to guide the model\cite{openai2024gpt4technicalreport} in selecting the answers.

In a straightforward approach, we employ vector retrieval for similar product search. Specifically, we utilize the general-purpose representation model BGE\cite{xiao2024c}, to encode products into vector embeddings. We then compute the cosine similarity between the query product and a candidate product pool to select similar items.Furthermore, to ensure the consistency of attribute items, we restrict product retrieval within the same category.
\subsection{Attribute Value Generation}
\label{sec2}
After completing attribute value retrieval and product retrieval, we integrate them into a unified template as training data. The combined prompt consists of five parts: task definition, note, similar products in the same category, product information, and candidate attribute values. The specific details are shown in the appendix\ref{sec:appendixA}.

We choose Qwen2.5\cite{qwen2025qwen25technicalreport} as the base model, leveraging their strong performance and robust open-source ecosystem. The training objective of the model is the classic next-word prediction task. Specifically, we exclude the loss computation of the prompt prefix, and instead focus on the special tokens and model output tokens.
$$L = - \sum_{t=1}^{T} \log P(y_t | y_{<t}, X)$$
Here, X represents the input to the model, y represents the generated token, t represents the position of the generation, and T represents the final generated sequence.
\section{Experiments}
\subsection{Data}
We evaluate our model on the large-scale product attribute recognition dataset Xianyu-PAVI. This dataset is sourced from the second-hand e-commerce platform Xianyu, where each product goes through a three-step process of model prediction, seller feedback, and manual review for attribute value assignment. The dataset contains 8,803 product categories, 26,645 category-attribute pairs, and 6.3 million category-attribute-attribute value triples. On average, each category has 3 attributes and 716 attribute values. The statistics of the dataset are shown in Tab \ref{tab:statistics_dataset}.

\subsection{Metrics}
Following TACLR\cite{su2025taclr}, we use micro-averaged precision@1, recall@1 and F1 score@1 to evaluate PAVI methods.For each attribute, the ground truth is a set of values V from the taxonomy. If the ground truth set is empty, a correct prediction (True Negative, TN) occurs when the model also predicts an empty set; otherwise, it is a False Positive (FP). When the ground truth set is not empty, the model’s top-1 output is a True Positive (TP) if it matches any ground truth value. Predicting an empty set in this case results in a False Negative (FN), while mismatched predictions are both False Positives (FP) and False Negatives (FN)\footnote{In previous work \cite{shinzato-etal-2023-unified}, evaluation metrics did not explicitly consider the false positive (FP) case, and instances involving both false positives (FP) and false negatives (FN) were categorized solely as false positives. In contrast, we employ more rigorous evaluation metrics that independently account for both FP and FN cases.}, as it simultaneously introduces an error and misses the correct value. Table 3 summarizes these outcomes. Final precision, recall, and F1 scores are computed by aggregating TP, FP, and FN counts across the dataset, providing a comprehensive performance evaluation.

In addition, for nonempty set of ground-truth attribute values, we introduce the Coverage metric to quantify the overlap between the retrieved attribute values and the ground-truth set; coverage is deemed achieved when the intersection of the two sets is nonempty.
\begin{table}
    \centering
    \caption{Statistics of the Xianyu-PAVI.}
    \label{tab:statistics_dataset}
    \begin{tabular}{lrrr}
    \hline
    {Statistic} & Train & Valid & Test  \\
    \hline
    \# Products       &   809,528 &  81,699 &  85,024 \\
    \# PA Pairs       & 3,584,462 & 358,582 & 458,954  \\
    \# Null Pairs     & 2,345,577 & 228,534 & 272,285 \\
    \hline
    \end{tabular}

\end{table}

\subsection{Baselines}
We compare our model with the one-stage PAVI methods, including classification, generation, and the current state-of-the-art retrieval-based methods:
\begin{itemize}
    \item \textbf{BERT-CLS} This model\cite{chen-etal-2022-extreme} treats PAVI as a multi-label classification task, where each attribute value is considered an independent label. It also performs masking on irrelevant attribute values based on the product category.
    \item \textbf{LLM} The basic LLM model treats PAVI as a generative task, learning the product attribute vocabulary during training. It then uses the product title description as input and the attribute-attribute value pairs as output.
    \item \textbf{Product-RAG} Building upon the LLM model, RAG recalls relevant products to provide more contextual information for attribute value identification.
    \item \textbf{TACLR} The current state-of-the-art PAVI model, which treats PAVI as a retrieval task. It combines contrastive learning and adaptive thresholding techniques to select attribute values by computing the similarity between the product description and the attribute values.
\end{itemize}

\subsection{Implementation Details}
\textbf{Retrieve Module}. For retrieving similar products, we employ BGE-base\cite{xiao2024c} as representation model and utilize cosine similarity as the similarity measurement. Furthermore, to ensure attribute consistency, retrieval is restricted to products within the same category. For retrieving attribute values, we employ the current state-of-the-art retrieval model, TACLR, to recall k attribute values.
\\
\textbf{Generate Module}. We trained our model based on Qwen2.5-7B-Instruct, using retrieved products of same category and retrieved values. We trained our model by full parameter fine-tuning, The core hyperparameters are as follows: 3 training epochs, batch size of 16, AdamW optimizer, maximumlearning rate of $2 * 10^{-5}$, 1\% warmup steps, cosine learning rate scheduler.

\section{Results}
\subsection{Main Result}

\begin{table*}
\centering
\caption{Performance comparison of classification, generation, and retrieval methods on Xianyu-PAVI}
\label{tab:main_res}
\begin{tabular}{l|ccccc}
\toprule
Methods& BERT-CLS & Qwen2.5 (Product-RAG) &Qwen2.5 (fine-tune) & TACLR& MVP-RAG(ours) \\
\midrule
pre & 50.9 & 58.3 &84.5 & 85.4& \textbf{93.8} \\
recall & 69.1&50.1 & 79.1& \textbf{87.1} & 85.3 \\
f1 & 50.5 & 63.2 & 81.7 & 86.2& \textbf{89.5} \\
\bottomrule
\end{tabular}
\end{table*}

Tab \ref{tab:main_res}. shows the comparative results of our model on the Xianyu-PAVI dataset. And MVP-RAG outperforms all the baselines, achieving the state-of-the-art F1-score on Xianyu-PAVI. We attribute the excellent performance of MVP-RAG to two advantages:
One is the retrieval-based candidate value generation which ensures the comprehensiveness of the attribute values, the other is incorporating similar products with corresponding attribute values from the same category as supplementary inputs to further improve the accuracy of the identification. The combination of these two strategies ensure both the completeness and accuracy of the attribute value recognition. Specifically, MVP-RAG outperforms the product-retrieval-based methods such as Qwen2.5(Product-RAG) by 26.3\% on F1-score and outperforms the attribute-retrieval-based methods like previous SOTA TACLR by 3.3\%. 

\subsection{Analysis}
\subsubsection{Impact of Product Counts}
\begin{figure}[t]
\centering
  \includegraphics[width=0.9\columnwidth]{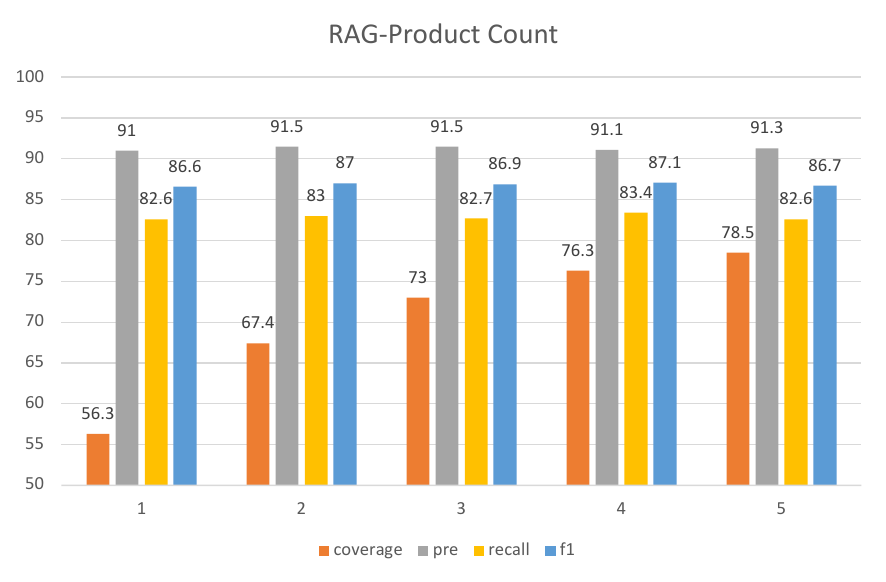}
  \caption{Performance analysis across Product Counts}
  \label{fig:rag-item}
\end{figure}
Fig \ref{fig:rag-item}. compares the impact on results as the number of retrieved products from the same category is varied. As the number of retrieved same-category products increases, the coverage of the test set's attribute values rises from 56.3\% to 78.5\% and gradually stabilizes. In contrast, the model's F1 score exhibits relatively minor fluctuations. This is because as the coverage increases with more similar product retrieval, attribute value generation becomes easier, leading to improved recall.
\subsubsection{Impact of Attribute Value Counts}
\begin{figure}[t]
\centering
  \includegraphics[width=0.95\columnwidth]{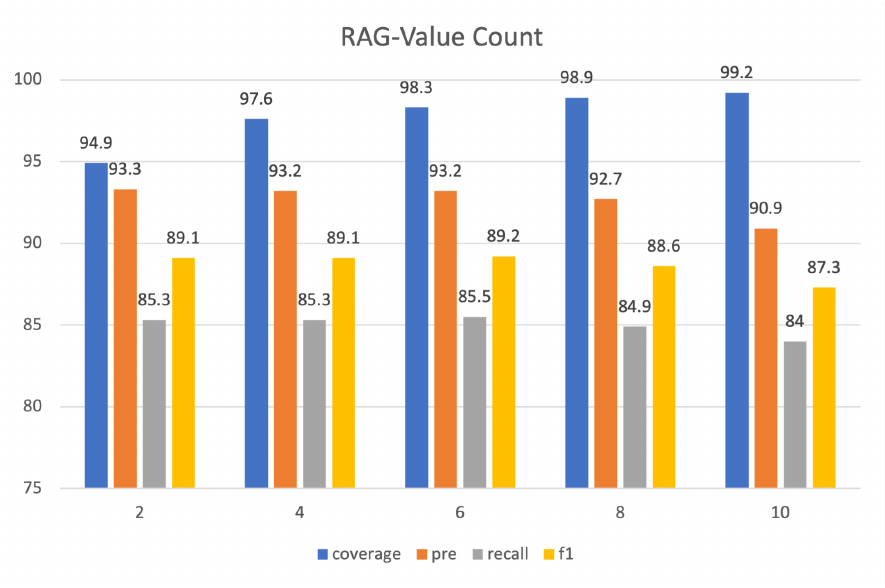}
  \caption{Performance analysis across Rag Value Counts}
  \label{fig:rag-value}
\end{figure}

Fig \ref{fig:rag-value}. compares the impact on results as the number of retrieved attribute values is varied. As the number of retrieved attribute values increases, the coverage of the test set's true attribute values gradually improves (94.9\% -> 99.6\%), approaching 100\%. The model's recall first increases and then decreases with the number of candidate attribute values, while precision declines overall. F1 and recall exhibit a similar pattern of initial increase followed by decrease, with the F1 score peaking at 89.2\% when the number of candidate values is 6. Clearly, as the number of candidate values increases, more information is introduced, with the initial information supplementation eventually leading to misinformation. Additionally, it is worth noting that compared to the TACLR retrieval model that provides candidate values, the model's precision improved by nearly 8\% (85.4\% -> 93.2\%), indicating that MVP-RAG is effective and leads to a significant enhancement in the online user experience.

\subsubsection{Accuracy Of Attribute Values And Products}
As the number of recalled attribute values increases, the coverage of true attribute values gradually increases. When the true result appears in the candidate attribute value, f1 can reach 92.6\%, which is 6.3\% higher than when it does not appear. At the same time, we control the number of recalled products to be 4, and analyze whether the accuracy of the attribute value of the recalled product has an impact on the results. For simple attributes with clear distinguishing meanings, such as brand and model, when the attribute value of the recalled product is 75\% wrong, MVP-RAG can still return the correct attribute value; when the error is higher than 75\%, MVP-RAG has a probability of returning an incorrect attribute value. At the same time, thinking models such as DeepSeek will first analyze whether the attribute value of the given retrieved product is wrong, and then make corrections to predict the attribute value. For attributes that do not have clear meanings, such as color function, MVP-RAG focuses on the product's own information and can still return the correct result when the attribute value of the recalled product is wrong. DeepSeek focuses on analyzing the information of the recalled product and is more inclined to be consistent with the attributes of the recalled product. The specific details are shown in the appendix\ref{Impact_of_incorrect_product_attribute_values}.

\section{Related Work}
\textbf{Product Attribute Value Identification} Product Attribute Value Identification (PAVI) is a critical task\cite{klemm2024importance} in e-commerce, with current methods broadly categorized into one-stage and two-stage paradigms. The two-stage approach first performs product attribute value extraction(PAVE)\cite{yang2022mave}, followed by standard value mapping, which can lead to cascading errors. The one-stage paradigm includes classification-based, generation-based, and retrieval-based methods. The classification-based approach\cite{fuchs-acriche-2022-product,chen-etal-2022-extreme} defines PAVI as a multi-label classification problem over a finite set of values, but suffers from a significant limitation in its inability to identify out-of-distribution (OOD) values, a critical shortcoming in the rapidly evolving e-commerce domain. Retrieval-based method\cite{su2025taclr} define PAVI as an attribute value retrieval task, but exhibit insensitivity to new attribute values and require redeployment to accommodate changes in the attribute library. Generation-based approaches cast PAVI as an end-to-end generation task, directly generating single or multi-attribute values from product descriptions, but face challenges with hallucination and output instability. The proposed model builds upon both generation-based and retrieval-based approaches, incorporating information from similar products to perform robust product attribute value identification.
\\
\textbf{Retrieval-Augmented Generation} Retrieval-based approaches\cite{leto2024optimalsearchretrievalrag} have been widely demonstrated to be highly effective for knowledge-intensive tasks, particularly in output-constrained settings, where the introduction of relevant knowledge can significantly mitigate the issue of hallucination. \cite{sabeh2024empiricalcomparisongenerativeapproaches} incorporated Retrieval-Augmented Generation (RAG) technologies in the PAVI task, utilizing a pre-trained T5\cite{raffel2023exploringlimitstransferlearning} to retrieve similar products and leverage the information to enhance the generation, outperforming direct attribute value generation methods by a substantial margin. In contrast, the unique contribution of current research lies in its utilization of both product-level and attribute-value-level information for the retrieval-augmentation process. By successfully applying this approach in the e-commerce domain, the proposed model is able to address the hallucination problem while also improving performance on the specific task at hand.

\section{Conclusion}
In this work, we present multi-value-product retrieval-augmented generation (MVP-RAG) method for the PAVI task. This approach addresses the limitations of existing PAVI methods, which often suffer from cascading errors, inability to handle out-of-distribution (OOD) attribute values, and lack of generalization capability. 

Comprehensive experiments on proprietary and public datasets demonstrated MVP-RAG's superiority over classification- and generation-based baselines. Notably, method achieved an F1 score of 89.5\% on the large-scale Xianyu-PAVI dataset. Beyond these experimental results, MVP-RAG has been successfully deployed on the real-world e-commerce platform Xianyu, processing millions of product listings daily and seamlessly adapting to dynamic attribute taxonomies, making it a practical solution for large-scale industrial applications.

\section*{Limitations}

MVP-RAG can accurately identify product attribute values, but there are still many limitations. First, only text information is used, and image or video information of the product has not been used. Multimodal input can provide information for attributes that are difficult to infer from text alone (such as color, material, or shape). Second, the reasoning output based on LLM takes a long time, and the model still needs to be accelerated and optimized in the future.

\bibliography{custom}

\appendix

\section{Example Appendix}
\label{sec:appendixA}
\textbf{Task Description}\\
Given the product description, category, similar same category product, attribute set, and candidate attribute values for each attribute, generate the attribute values of the product. Unrecognizable attributes can be returned as unknown.\\
\textbf{Note}\\
1. The attribute value of the product does not necessarily appear in the reference product attribute value and candidate attribute value\\
2. If the attribute value of the product exists but does not appear in the reference product attribute value and candidate attribute value, it can be generated, but it is best not to do so\\
3. If the attribute value of the product does not exist, return \textit{None}\\
4. The probability of the given candidate attribute value decreases from the front to the back\\
\textbf{Reference product information}\\
Product description: 2 Sony E-mount Tamron back covers, 1 body front cover. New Year's sundries! \\
Category: SLR body cover\\
Attribute value:\\
Brand: Sony\\Condition: None\\
Product description: Canon 60d back cover set button set button\\ 
Category: SLR body cover\\
Attribute value:\\
Brand: Canon\\
Condition: None\\
\textbf{Product information}\\
Product description: Nikon D series SLR camera front and rear covers, dust cover, body cover, suitable for Nikon SLR camera D810D7100/D610/D3200/D5100D800 body cover, lens rear cover \\
Category: SLR body cover \\
Attribute collection: brand, condition\\
\textbf{Attribute candidate value} \\
Brand: Nikon, None\\
Condition: None, slight signs of use\\
Brand:Nikon\\
Condition:None\\

\section{Impact of incorrect product attribute values}
\label{Impact_of_incorrect_product_attribute_values}
\textbf{Product.}

Chanel new sale S engraved 19 black gold small size, how hard to buy not to say, no bargaining, only inspection treasure\\
\textbf{Similar same category Product.}\\
Chanel 20a Haute Couture, size 36 black short jacket with gold buttons, the same style as Zhou Xun, basically brand new, very good-looking, authentic guaranteed, supports identification!\\
Chanel four-pocket black jacket, a classic among Chanel classics, size 34, 99\% new, purchased in a domestic counter, because it is an old model, it cannot be inspected, and those who can distinguish the authenticity can come, no bargaining, out of print collection, suitable for thin girls under 100 pounds, SF Express cash on delivery, no refund or exchange.\\
23 autumn and winter Chanel mer select Italy jin import black navy blue glitter silk 50 wool small fragrance care products anonymous eyiu bought size s if it is not suitable after arrival, rush to sell it. The goods have been received and sold.\\
\[Ready Stock\] Chanel 23a black denim cc jacket size 36 Brand new full set, direct style can be sent for free inspection within 3-5 working days, direct style delivery, serious buyers only.\\
\textbf{Explicit Attribute-Brand}\\
\textbf{MVP-RAG}\\
When the brand of the given three products is changed to Dior, the result returned is always Chanel. However, when the brand of all four products is changed to Dior, the result returned is Dior.\\
\textbf{DeepSeek}\\
\textbf{rethink}\\
First, there are four examples in the reference product information provided by the user. Although "Chanel" is mentioned in the product description, the brand in the attribute value is marked as "Dior", which is obviously inconsistent. This may be an error or a test case. However, according to the task description, I need to generate attribute values based on the given information and return unknown when it cannot be recognized.
The description of the target product is "Chanel New Sale S Engraved 19 Black Gold Small", which clearly mentions "Chanel", so the brand should be "chanel". But the brands in the reference examples are all Dior. Here we need to judge whether to follow the error in the reference example or fill in the brand correctly according to the product description. According to the second point of the task description, if the attribute value exists but does not appear in the reference, it can be generated but it is better not to. But here the brand is clearly in the description, so it should be filled in correctly instead of following the wrong example.\\
\textbf{Answer}\\
No matter how many wrong results are given, it can always return to Chanel.\\
\textbf{Explicit Attribute-Quality}\\
\textbf{MVP-RAG}\\
Always returns a new attribute value that is consistent with the correct result.\\
\textbf{DeepSeek}\\
\textbf{rethink}\\
Let’s first look at the information of the reference products. The description of the first product mentions “basically new”, and the condition in the attributes is “almost new”. The description of the second product says “99\% new”, and the condition is also “almost new”. The descriptions of the third and fourth products mention “out” and “brand new complete set” respectively, and the condition is also “almost new”. It seems that no matter whether the description is “basically new”, “99\% new” or “brand new”, the condition is indicated by “almost new” in the attributes. Is this because in the second-hand or resale market, these descriptions are all classified as almost new, rather than brand new? Or it may be a unified processing method in the task, so even if the description says “brand new”, the attributes still use “almost new”.\\
\textbf{Answer}\\
When almost new appears, DeepSeek will tend to give almost new results.
\end{document}